\documentclass[preprint,10pt]{elsarticle}

\usepackage{amssymb}
\usepackage{amsmath}

\usepackage{amsfonts}
\usepackage{algorithm}
\usepackage{array}
\usepackage{subfig}
\usepackage{textcomp}
\usepackage{stfloats}
\usepackage{url}
\usepackage{verbatim}
\usepackage{graphicx}
\usepackage{cite}

\usepackage{appendix}

\usepackage{gensymb}

\let\Require\relax
\let\Ensure\relax
\usepackage{algpseudocode}

\usepackage{xcolor}
\usepackage{pifont}
\newcommand{\cmark}{\ding{51}}

\usepackage{multirow}
\usepackage{multicol}

\usepackage{hyperref}
\hypersetup{
    colorlinks=true,
    linkcolor=blue,
    citecolor=blue,
    urlcolor=blue,
    filecolor=magenta,
    pdftitle={Overleaf Example},
    pdfpagemode=FullScreen,
}

\usepackage{orcidlink}

\journal{Journal of Non-Crystalline Solids}

\begin{document}

\begin{frontmatter}

\title{Enhancing Reconstruction Capability of Wavelet Transform Amorphous Radial Distribution Function via Machine Learning Assisted Parameter Tuning}

\author[lab1]{Deriyan Senjaya}
\author[lab2]{Stephen Ekaputra Limantoro}

\affiliation[lab1]{
  organization={Department of Physics, National Tsing Hua University},
  city={Hsinchu},
  postcode={30013},
  country={Taiwan}
}

\affiliation[lab2]{
  organization={Department of Electrical Engineering and Computer Science,
                National Yang Ming Chiao Tung University},
  city={Hsinchu},
  postcode={30010},
  country={Taiwan}
}

\begin{abstract}

Understanding atomic structures is crucial, yet amorphous materials remain challenging due to their irregular and non-periodic nature. The Wavelet Transform Radial Distribution Function (WT-RDF) offers a physics-based framework for analyzing amorphous structures, reliably reconstructing the first and second Radial Distribution Function (RDF) peaks and overall curve trends in both binary $(\text{Ge}_{0.25}\text{Se}_{0.75})$ and ternary $\text{Ag}_{x}(\text{Ge}_{0.25}\text{Se}_{0.75})_{100-x}$ ($x=5,10,15,20,25$) systems. Despite these strengths, WT-RDF shows limitations in amplitude accuracy, which affects quantitative analyses such as coordination numbers. The shortcoming arises from improper parameter ($a$, $b$, $K_{f}$, $\tilde{C}$, and $\Lambda$) selection, as the parameters intrinsically represent atomic interactions within amorphous materials. This study addresses the issue by optimizing WT-RDF parameters using a machine learning approach via learnable parameter optimization, parameter bounding, and selective loss, producing the enhanced WT-RDF$+$ framework. WT-RDF$+$ improves the precision of peak reconstructions and outperforms benchmark Machine Learning (ML) models, including Radial Basis Function (RBF) and Long Short-term Memory (LSTM), when trained on only 25$\%$ of the binary dataset. Specifically, the machine learning benchmarks are defined as regressors with radial distance $r$ input and $G(r)$ output taken from Ab Initio Molecular Dynamics (AIMD) RDF simulation, not the reduced structure factor $S_R(q)$ to $G(r)$ inversions. These results demonstrate that WT-RDF$+$ is a robust and reliable model for RDF reconstruction of Ge–Se and Ag-Ge-Se family.

\end{abstract}


\end{frontmatter}

\section{Introduction}
\label{sec1}
\begin{figure}[ht]
\centering
\includegraphics[scale=0.4]{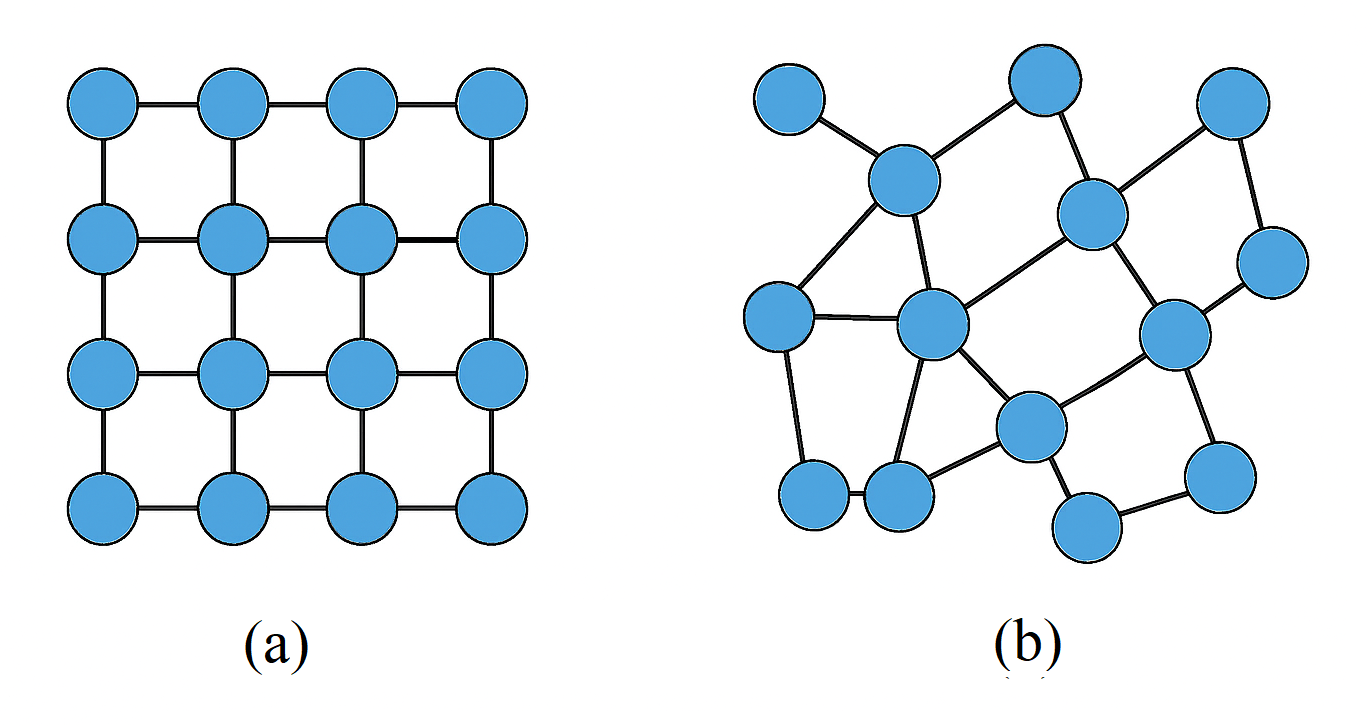}
\caption{Illustration of the atomic structure of a material with (a) crystalline structure and (b) amorphous structure}
\label{fig:fig1}
\end{figure}

In general, material atomic structures fall into two categories, \textit{i.e.}, crystalline and amorphous \citep{callister}. Specifically, atoms in crystalline materials are arranged in a highly ordered, periodic structure, such as iron with a body-centered or face-centered cubic structure and table salt (NaCl) \citep{callister, richard}. In contrast, amorphous materials lack long-range order due to their disordered structure, with a common example being glasses \citep{zbigniew, sati2016}. Illustration of both structural types is displayed in Fig. \ref{fig:fig1}. The key differences between those materials make amorphous structures challenging to characterize and analyze \citep{Madanchi2024, Alec2023}. This is clear from the available tools for determining their atomic arrangements. Crystalline structures can be identified using X-ray diffraction and analyzed with Bragg’s law, $2d \sin\theta=n\lambda$, where $d$ is the interatomic spacing, $\theta$ the diffraction angle, and $\lambda$ the X-ray or neutron wavelength \citep{callister}. However, Bragg’s law does not apply to amorphous materials, making physicists typically rely on the Radial Distribution Function (RDF) when studying amorphous structures \citep{zallen, zbigniew, ziman, elliot}.

The RDF, rooted in classical statistical mechanics, describes the number of atoms located at a given radial distance from a chosen reference atom \citep{zallen, zbigniew, senjaya2020, Petkov1989, zhao2010, shevchenko2019}. If an amorphous material has a radial density distribution $\rho(r)$, then the RDF $g(r)$ can be derived using Eq. (\ref{1}). Experimentally, it can also be obtained by Fourier transforming the reduced structure factor $S_{R}(q)=q(S(q)-1)$ from X-ray diffraction data, yielding the reduced RDF in Eq. (\ref{2}) \citep{petkov, Petkov2005, Petkov1989, senjaya2020}.

\begin{equation}
    \label{1}
    g(r)=\frac{dN(r)}{dr}=4\pi r^{2}\rho(r)
\end{equation}

\begin{equation}
    \label{2}
    G(r)=\frac{g(r)}{r}=\frac{2}{\pi}\int^{q_{\text{max}}}_{q_{\text{min}}}S_{R}(q) \sin (q r) dq
\end{equation}

The characteristic output of Eq. (\ref{1}) is a series of peaks at specific radial distances $r$. The first peak (FP) corresponds to the distribution of atoms located at the nearest-neighbor distance from the chosen reference atom \citep{senjaya2020}. The second peak (SP) reflects the distribution of second-nearest neighbors, and subsequent peaks follow the same interpretation \citep{senjaya2020}.

Additional issues arise when using Eq. (\ref{2}). If the X-ray source has a low wave vector ($q < 8\ \text{\AA}^{-1}$), such as Cu K-$\alpha$ and Mo K-$\alpha$, the resulting RDF has poor resolution. Consequently, the peaks that describe the atomic distribution in the amorphous material become indistinct and difficult to interpret. To address this limitation, a new mathematical approach is required. As demonstrated in \citep{senjaya2020}, the Wavelet Transform (WT) offers an effective alternative. The WT-based RDF formulation used in \citep{senjaya2020} is shown in Eq. (\ref{3}).

\begin{equation}
    \label{3}
    G(r)=\Lambda\int^{q_{\text{max}}}_{q_{\text{min}}}S_{R}D^{*}\left(\frac{qr-b}{a}\right)dq
\end{equation}
Here, $D$ is a wavelet function introduced in \citep{senjaya2020}, constructed using the mean-field method (discussed further in Sec. \ref{II}). The explicit form of $D$
is given in Eq. (\ref{4}). For more details about $\Lambda$, $\beta_{s,p}$ and $\Phi$, please refer to Sec. \ref{II}. 

\begin{equation}
    \label{4}
      D=\sum^{\infty}_{s=0} \beta_{s,p} \left( \dfrac{qr-b}{2a} \right )^{2s+p}
    + K_{f} \Phi \left( \dfrac{qr-b}{a} \right ) + \tilde{C}
\end{equation}

To determine the parameters $a$, $b$, $K_{f}$, $\tilde{C}$, and $\Lambda$ \citep{senjaya2020} used experimental $S_{R}(q)$ data from the amorphous Ge–Se sample $\text{Ge}_{0.25}\text{Se}_{0.75}$ (referred to as the binary system), obtained via X-ray diffraction using a Cu K-$\alpha$ source ($\lambda\approx 1.54 \ \text{\AA}$) and transformed using Eq. (\ref{3}). The initial parameters were first chosen arbitrarily and then manually tuned to match the alignment of WT-derived RDF and the Ab Initio Molecular Dynamics (AIMD) RDF simulation. The best reconstruction result of WT-RDF was achieved with $a=0.590$, $b=4.600$, $K_{f}=0.106$, $\tilde{C}=0.212$, and $\Lambda=-4.700$. These optimized parameters were then tested on Ag-doped Ge–Se amorphous materials $\text{Ag}_{x}(\text{Ge}_{0.25}\text{Se}_{0.75})_{100-x}$ with $x=5,10,15,20,25$ (the ternary systems). The WT-RDF remained consistent, accurately identifying the first and second peak positions with errors between 2.44\% and 10.16\%, while also preserving the overall RDF trend \citep{senjaya2020}.

However, the RDF amplitudes obtained using the WT approach still do not fully match the AIMD RDF for each material \citep{senjaya2020}. This limitation makes the WT-based RDF model insufficient for accurately determining amorphous structures. In particular, it affects the calculation of coordination numbers (CN), the number of nearest-neighbor atoms around a chosen central atom, which depends strongly on the RDF amplitude, as indicated in Eq. (\ref{5}) \citep{ziman}.

\begin{equation}
    \label{5}
    CN=\int_{\text{first peak}}4\pi r^{2}g(r) dr
\end{equation}

Therefore, it was still unclear if the WT-based mathematical model was defective or if the previously selected parameters were precise.  In order to address this, \citep{senjaya2023} used the Born-Mayer (BM) potential to evaluate the WT-RDF model (Eq. (\ref{3})) against RDFs derived using classical density functional theory (CDFT). Because it adequately captures short-range atomic interactions, such as ionic bonds, which are typical of amorphous materials, the BM potential was chosen \citep{Guillaume2014}. The WT-RDF reliably reproduces the coefficients of the BM potential, which expresses atomic repulsion and attraction, according to results from \citep{senjaya2023}.  This validates the WT-RDF model from \citep{senjaya2020} and shows that inaccurate parameter selection, rather than model defects, is the main cause of disparities in RDF amplitude.

To find the RDF profile of amorphous materials, researchers use simulation \citep{rosset,ctu} and reconstruction methods \citep{senjaya2020} using experimental data. In simulations, the methods often used are molecular dynamics \citep{rosset} and density functional theory \citep{ctu}. The simulation approach often involves many atoms to achieve accurate RDF profile results, while the use of reconstruction is faster because, with minimal data, the RDF profile can be known, and physical quantities related to the structure of amorphous materials, such as interatomic distances, coordination numbers, and ring statistics, can be obtained efficiently. Here, the WT-RDF method shows good reconstruction capabilities. While WT-RDF has shown promising results, it is currently less established compared to standard RDF workflows \citep{petkov}. Therefore, WT-RDF, which focused on calibrated reconstruction within the Ge-Se family, is chosen as the baseline and extended in this study.

Inspired by \citep{senjaya2023}, which demonstrated that just the model parameters need to be chosen more precisely, this work suggests utilizing machine learning (ML) to tune the WT-RDF model parameters. It was selected because ML uses the calculus of variations to improve parameters for reconstructive models \citep{Deisenroth}.  Furthermore, substantial progress in component designing and production has been made possible by machine learning \citep{manufacturing_solder, manufacturing_signal}. WT-RDF$+$ is an improved WT-RDF model with ML-optimized parameters. Two machine learning models, the Radial Basis Function (RBF) and Long Short-Term Memory (LSTM), are used to benchmark the performance of WT-RDF$+$. Finally, the main research contributions are summarized into two-fold:

\begin{enumerate}
    \item To characterize the structure of amorphous materials, we present WT-RDF$+$, a physics-based model improved using machine learning. To effectively optimize the model, parameter bounding and a selective loss function are introduced. According to experimental findings, WT-RDF$+$ performs better than the original WT-RDF model on a variety of datasets and data ratios.
    \item Facing the training data scarcity, the performance of data-driven ML models (RBF and LSTM) significantly declines. On the other hand, WT-RDF$+$, a physics-based model, is resilient even in the presence of sparse data because it is grounded in mathematical formulations and physical principles. This demonstrates the reliability and generality of WT-RDF$+$ despite the limited availability of training data.
\end{enumerate}

\section{Wavelet Transform Amorphous RDF}
\label{II}
A mathematical framework called WT-RDF uses the Wavelet Transform (WT) to calculate the RDF of amorphous materials \citep{senjaya2020}. WT was selected primarily for two reasons. In order to overcome the poor resolution of Fourier Transform (FT) when analyzing $S_{R}(q)$ data from low-energy X-rays like Cu K-$\alpha$, it first functions as a "mathematical microscope," offering detailed resolution in local regions of a function through its dilation ($a$) and translation ($b$) parameters \citep{nanavati2004}. Second, as demonstrated in Eq. (\ref{6}) \citep{bopardikar1998}, WT permits the flexible construction of wavelet functions $f(x)$ as long as they meet the usual WT criteria, which include the zero-mean and orthogonality constraints.

\begin{equation}
    \label{6}
    \int^{\infty}_{-\infty} x|f(x)|^{2}dx=0  \ ;  \int^{\infty}_{-\infty}f_{i}^*(x)f_{j}(x)dx=\delta_{ij}
\end{equation}

\subsection{Formulation of Amorphous WT-RDF}
\begin{figure}[ht]
\centering
\includegraphics[scale=0.42]{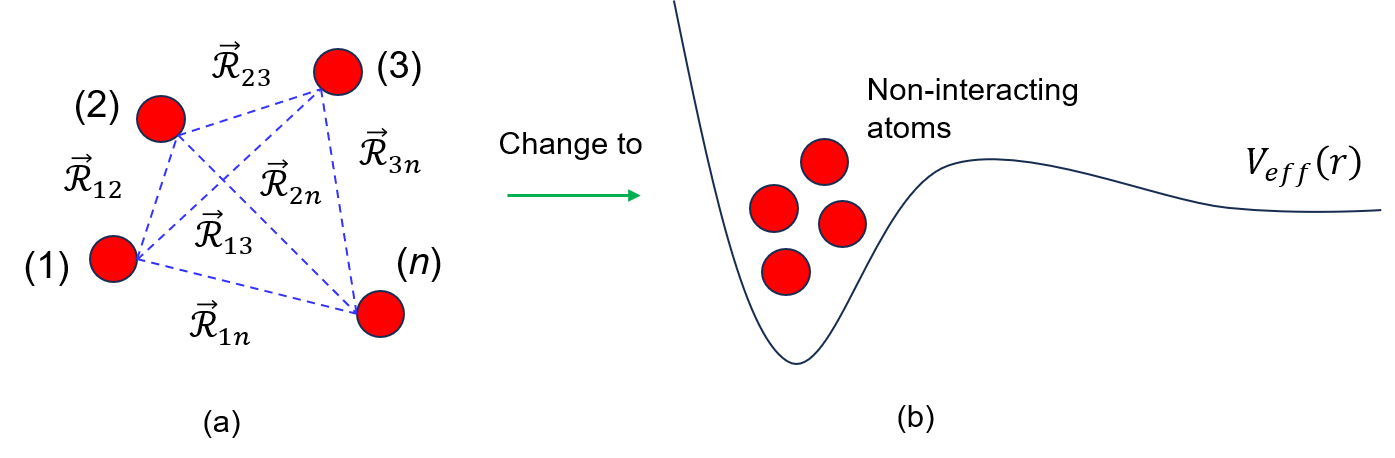}
\caption{Illustration of the mean field method that is adapted in the WT-RDF formulation}
\label{fig:fig2}
\end{figure}

Think of an amorphous substance as a collection of atoms that are bound together (Fig. \ref{fig:fig2}).  Because each atomic nucleus has three coordinates (x, y, and z), solving the Schrödinger equation for all atomic nuclei is analytically difficult and results in a 3N-dimensional problem.  The mean-field approach, which is based on the Kohn-Sham Density Functional Theory, is used to simplify this.  As seen in Fig. \ref{fig:fig2}, it assumes that atoms do not directly interact and that the consequences of interactions like spin, exchange, and correlations are represented by an effective potential $V_{\text{eff}}(r)$.

To simplify $V_{\text{eff}}(r)$, it is first approximated as a one-dimensional box potential with depth $V_{o}$ and length $L$\citep{senjaya2020}. Solving the Schrodinger equation for a single atom in this potential yields the wave function $\psi_{n}(r)=A_{n}\sin (k_{n}r)$ for $0<r<L$ \citep{senjaya2020}. For a system of $N$ atoms, the total wave function is $\Psi(r)=\sum_{n}c_{n}\psi_{n}(r)$, giving a density $\rho(r)=\sum_{n}f_{n}|\psi_{n}(r)|^{2}$ , with $f_{n}=c_{n}^{*}c_{n}$ as the occupation coefficient \citep{senjaya2020}. Since $\rho(r)$ approximates a hypergeometric function related to the first-kind Bessel function when $f_{n}, A_{n}$, and $k_{n}$ are determined \citep{abramowitz1965}, the wavelet function in Eq. (\ref{6}) must incorporate the first-kind Bessel function. This term, shown in Eq. (\ref{7}), includes $\beta_{s,p}$, the linear combination coefficient, and $p$, the function order.

\begin{equation}
    \label{7}
    D_{\text{known}}=\sum_{s=0}^{\infty}\beta_{s,p}\left(\frac{qr-b}{2a}\right)^{2s+p}
\end{equation}

To refine the wavelet function in Eq. (\ref{7}), an additional “unknown” term is added (Eq. (\ref{8})), with $\Phi$ defined in Eq. (\ref{9}) to capture short-range atomic interactions \citep{senjaya2020}. Combining Eqs. (\ref{7}) and (\ref{8}) forms the wavelet function $D$ in Eq. (\ref{4}). To satisfy wavelet conditions (Eq. (\ref{6})), $p$ must be even and $\Phi$ an even function \citep{senjaya2020}. Once $D$ meets these criteria, it is substituted into Eq. (\ref{3}) to obtain the WT-RDF.

\begin{equation}
    \label{8}
    D_{\text{unknown}}=\Phi\left(\frac{qr-b}{a}\right)+\tilde{C}
\end{equation}
\begin{equation}
\label{9}
    \Phi=\left(\frac{qr-b}{a}\right)^{4}\exp\left[{-\left(\frac{qr-b}{a}\right)^{2}}\right] 
\end{equation}

\subsection{Application of Amorphous WT-RDF to Ge-Se (Binary) \& Ag-Ge-Se (Ternaries) based amorphous solids}
\begin{figure}[ht]
\centering
\subfloat[\centering]{{\includegraphics[width=6cm]{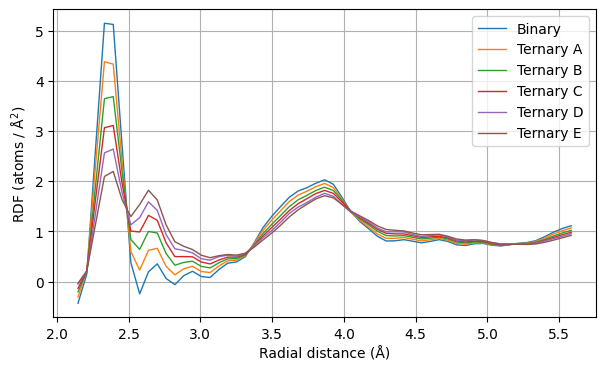}}}
\subfloat[\centering]{{\includegraphics[width=6cm]{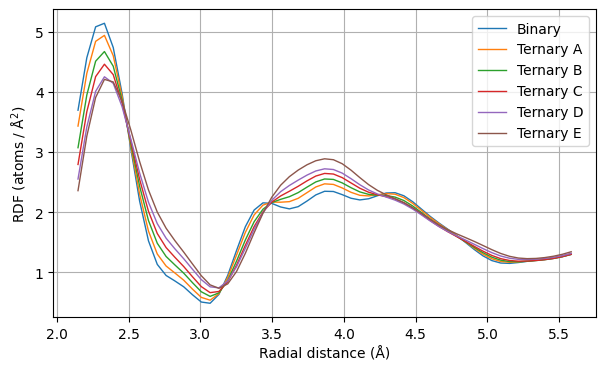}}}
\caption{Comparison between (a) AIMD RDF, and (b) WT-RDF for Ge$_{0.25}$Se$_{0.75}$ (binary) and Ag$_x$(Ge$_{0.25}$Se$_{0.75}$)$_{100-x}$ ($x$ = 5, 10, 15, 20, 25) (ternary A to E). The parameters are $a$ = 0.590; $b$ = 4.600; $K_f$ = 0.106; $\Lambda$ = $-$4.700; $\tilde{C}$ = 0.212.}
\label{fig:fig4}
\end{figure}

As mentioned in Section I, first, the existing parameters, namely $a$, $b$, $K_{f}$, $\tilde{C}$, and $\Lambda$, are determined arbitrarily. Then, manual tuning is performed on the binary data and the value $a=0.590$, $b=4.600$, $K_{f}=0.106$, $\tilde{C}=0.212$, and $\Lambda=-4.700$.  is obtained. This value is then applied to the ternary data, and the consistency of the reconstruction of the distances between the first and second nearest neighbors is obtained, as shown in Fig. \ref{fig:fig4}.

\begin{figure}[ht]
\centering
\includegraphics[scale=0.4]{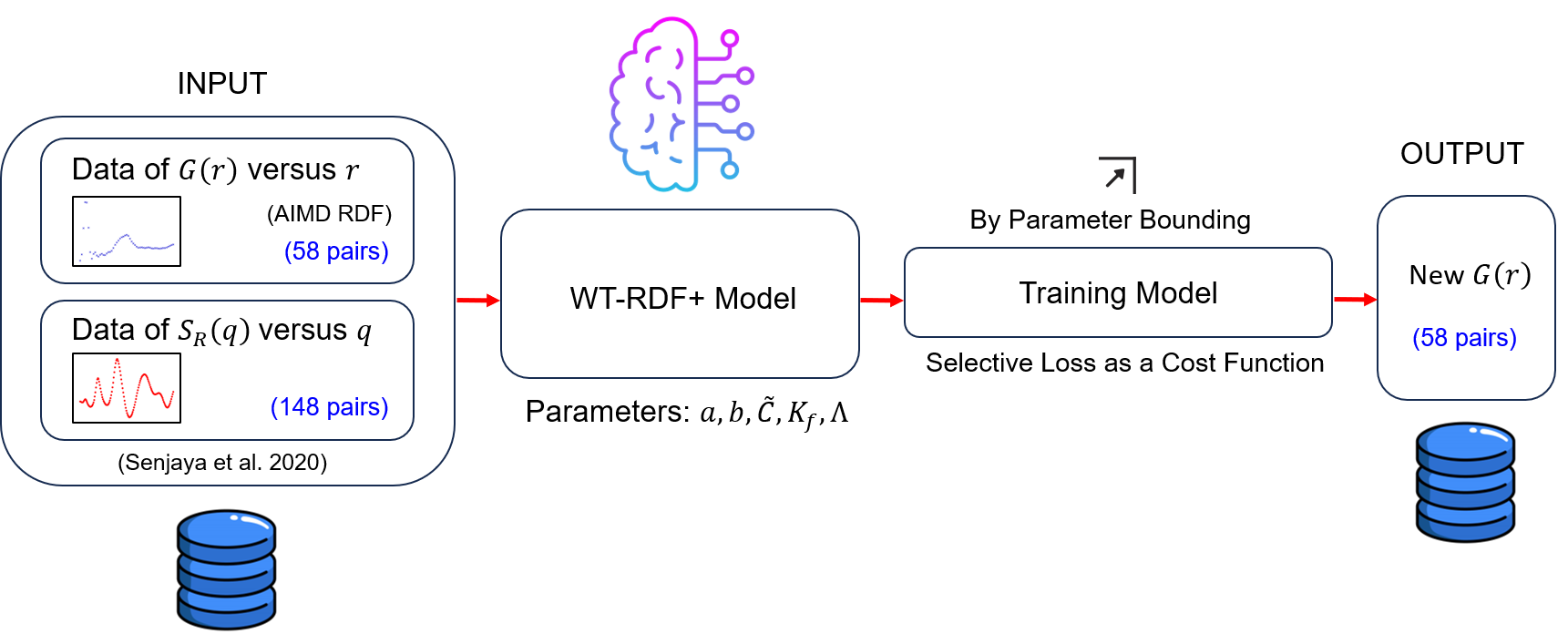}
\caption{The workflow of the proposed WT-RDF$+$. $r$ data and $S_{R}(q)$ versus $q$ data are used as training input to optimize the model with $a,b,\tilde{C},K_{f}$, and $\Lambda$ parameters. In the training phase, the $G(r)$ (AIMD data) is used for the loss function, minimizing the discrepancy with the new $G(r)$ (reconstruction output). The parameter bounding and selective loss were used as cost functions for training. In the inference phase, $r$ data and $S_{R}(q)$ versus $q$ data are required as the input of the trained WT-RDF$+$ model, resulting in the new $G(r)$.}
\label{fig:fig5}
\end{figure}

\section{Method}
In this research, we employ the regression method using machine learning models (\textit{i.e.}, Radial Basis Function (RBF) and Long Short-Term Memory (LSTM)) for benchmarking. Compared with Fourier Transform reconstruction results for the same $q$-range of WT-RDF studied in \citep{senjaya2023}, regression-based machine learning models achieve accurate RDF reconstruction. Therefore, machine learning is used as a benchmark in this study. We compare the best reconstruction results of the machine learning techniques with those of our proposed WT-RDF$+$. Note that the machine learning benchmark models are regressors of $r$ and $G(r)$, not scattering to RDF inversion models. Further, the workflow of the proposed WT-RDF$+$ is depicted in Fig. \ref{fig:fig5}. Specifically, the training input data, including $r$ data and data of $q$ and $S_R$($q$), is fed to the model. Moreover, the parameters are optimized using the proposed parameter bounding and selective loss to improve model reconstruction. The gap between $G(r)$ of AIMD data and the reconstructed $G(r)$ is minimized using the selective loss function. In the inference phase, the $r$ data and data of $q$ and $S_R$($q$) are fed to the trained model, resulting in the new $G(r)$.

\subsection{Machine learning benchmark}
We employ two types of machine learning models for evaluating the reconstruction of radial distribution data, including the Radial Basis Function (RBF) \citep{rbf} and Long Short-Term Memory (LSTM) \citep{lstm} models. Specifically, RBF is a feed-forward network for function approximation, while LSTM is an effective model to handle sequential data, including non-linear functions \citep{lstm2}. With these two distinct network characteristics, a comprehensive benchmark is established for comparative analysis against our physical model, WT-RDF$+$.

\begin{table*}[!ht]
\renewcommand{\arraystretch}{1.3}
\caption{Grid search on the reconstruction of the RBF network on binary data. The best results are highlighted in \textbf{bold}.}
\label{table_rbf_grid}
\centering
\resizebox{\textwidth}{!}{
\begin{tabular}{c|cccccccccccccccc}
\bfseries{$M$} & 16 & 32 & 64 & 16 & 16
& 32 & 64 & 16 & 64 & 128 & 256 
& 128 & 256 & 512 & 512 & 1024
\\
\bfseries{$\beta$} & 16 & 16 & 16 & 32 & 64
& 64 & 64 & 128 & 256 & 256 & 256
& 512 & 512 & 512 & 1024 & 1024
\\
\hline
\bfseries{MAE $\downarrow$} & 0.930 & 0.918 & 0.932 & 0.898 & 0.683
& 0.582 & 0.585 & 0.346 & 0.193 & 0.182 & 0.234
& 0.128 & 0.082 & 0.078 & \textbf{0.029} & 0.050
\\
\end{tabular}
}
\end{table*}

\subsubsection{Radial basis function network}

\begin{equation}
\phi_j(r) = \exp\Big(-\beta \| r - \mu_j \|^2 \Big), \quad j=1,\dots,M
\end{equation}

\begin{equation}
\hat{y} = \sum_{j=1}^{M} w_j \phi_j + b
\end{equation}

where $r$ is the input, $\beta$ is the width parameter to control the spread of the Gaussian function, $\hat{y}$ is the model output, $M$ is the number of RBF neurons, $w_j$ is the linear weights, $\mu_j$ is the center of the $j$-th, and $b$ is the bias term. 

To achieve the best results, we conduct a grid search over two parameters: $\beta$ and $M$. As shown in Table \ref{table_rbf_grid}, we can see that $\beta$ has a major influence on lowering the Mean Absolute Error (MAE). Based on the result, we use $M$ of 512 and $\beta$ of 1024 for comparison.

\begin{table}[!ht]
\renewcommand{\arraystretch}{1.3}
\caption{Grid search on the reconstruction of the LSTM network on binary data. The best results are highlighted in \textbf{bold}.}
\label{table_lstm_grid}
\centering
\begin{tabular}{c|cccccc}
\bfseries{$h_t$} & 8 & 16 & 32 & 64 & 128 & 256
\\
\hline
\bfseries{MAE $\downarrow$} & 0.069 & 0.033 & 0.036 & \textbf{0.023} & 0.026 & 0.035
\\
\end{tabular}
\end{table}

\subsubsection{Long short-term memory}

\begin{equation}
\begin{aligned}
h_t, c_t &= \text{LSTM}(r_t, h_{t-1}, c_{t-1}),\\
\hat{y_t} &= w h_t + b
\end{aligned}
\label{eq_6}
\end{equation}

The LSTM network is designed with a hidden state $h_t$ to capture temporal dependencies in the input sequence. Given the input data $r_t$, the hidden state $h_t$, and cell state $c_t$ are updated at each time step $t$, as shown in Eq. (\ref{eq_6}). The output of the hidden layer, $h_t$, is then passed through a fully connected linear layer to produce the final output $\hat{y_t}$. This configuration enables the model to effectively learn sequential patterns while maintaining a moderate level of model complexity.

We conduct another grid search on $h_t$ in Table \ref{table_lstm_grid}. Based on the result, we observe that LSTM, a time series model, provides an effective approach for modeling radial distribution function data. We further employ an LSTM network with a hidden state comprising 64 units.

\subsection{\texorpdfstring{WT-RDF$+$}{WT-RDF+}}
To improve the reconstructive performance of a physics-based model with machine learning assistance, three key factors have to be considered. Those are parameter optimization, parameter bounding, and selective loss. Model learning is driven primarily by optimizing its parameters. However, solely optimizing the parameter might lead the model to get stuck in a suboptimal region of the loss surface. Therefore, a parameter bounding method is adopted to ensure that the model converges to a better solution and avoids exploding gradients where weights change drastically. Moreover, a selective loss is incorporated to guide the model toward better reconstructive ability on peaks.

\subsubsection{Parameter optimization}
We converted the static parameters in our physical formula, denoted by $\theta$, into learnable parameters to achieve optimal results.

\begin{equation}
\begin{aligned}
    \hat{y} = \Lambda \int_{q_{min}}^{q_{max}} S_R(q) 
    \biggr[ \sum^{\infty}_{s=0} \beta_{s,p} \left( \dfrac{qr-b}{a} \right )^{2s+p} + \Gamma \biggr],
    \\
    \text{where } \Gamma = K_{f} \Phi \left( \dfrac{qr-b}{a} \right ) + \tilde{C}  
\label{eq:parameter_optimization}
\end{aligned}
\end{equation}

\begin{equation}
\theta = \{a, b, K_f, \tilde{C}, \Lambda \}, \quad \theta \text{ are learnable parameters.}
\end{equation}

where $\hat{y}$ refers to $G(r)$, which is the model reconstruction of input data $r$. More information about Eq. (\ref{eq:parameter_optimization}) can be found in Section I. 

For a clearer presentation, we further discuss the role and effects of each learnable parameter from a mathematical perspective in the following:
\begin{itemize}
    \item $a$ is a dilation coefficient controlling the scaling of $qr-b$. Smaller values of $a$ lead to a larger magnitude of the outputs, making the reconstruction flatter. On the contrary, higher values result in a reconstruction with more fluctuations.
    \item $b$ is a translation coefficient shifting the function $qr$, making the reconstruction slightly move horizontally and transforming the magnitude of the wave. Additionally, the peaks are out of reach due to extreme values. 
    \item $K_f$, coupling parameter that controls the interaction strength among the atoms. This parameter affects the reconstruction vertically and slightly changes the width of the valley or first minimum.
    \item $\tilde{C}$ suggests a bias/intercept to shift the output vertically.
    \item $\Lambda$ is used for normalization, scaling the entire result of the integral and summation. This parameter controls the height of the peaks. Larger values result in shorter peaks, and vice versa. 
\end{itemize}

This discussion will help in verifying whether the optimized parameter aligns with each role.

\subsubsection{Parameter bounding}
Making a constraint on the parameter is important to avoid vanishing and exploding gradients \citep{parameterbounding}. This technique may seem simple, but it is crucial for our physics-based model, particularly for sensitive parameters. We clip the parameters with the following equation:

\begin{equation}
\theta^{(i)} = \min\big(\max(\theta^{(i-1)}, n_{2}^{(i)}), n_{1}^{(i)}\big)
\end{equation}

where $n_{2}$ is the lower bound, $n_{1}$ is the upper bound for each iteration $i = 1,2,..., I$. To find the optimal bounding values, we use small intervals centered on the initial value and execute this process iteratively in the $i$-th step. 

We conduct an experiment and find that the parameter $a$ is highly sensitive compared to the other parameters due to its scaling factors. In the last iteration, we bound the parameter $a$ with $n_{2}$ = 0.600 and $n_{1}$ = 0.610. To further improve the performance, we also bound the parameter $K_f$, which is less sensitive than $a$, with $n_{2}$ = 0.010 and $n_{1}$ = 0.300. The rest parameters are updated gradually without clipping.

\subsubsection{Selective loss}
We introduce selective loss for WT-RDF to guide the model during training so that it prioritizes minimizing the errors associated with the peaks instead of global MAE. 

\begin{equation}
\mathcal{L}_{SL} = \sum^{P}_{p=1} \sum^{N}_{n=1} m_{n,p} | y_{n,p} - \hat{y}_{n,p} | 
\end{equation}

where $N$ is the number of samples, $P$ is the number of peaks (two peaks in this case),  $y_{n,p}$ is the ground-truth for $n$-th sample at the $p$-th selected peak bin, $\hat{y}_{n,p}$ is the model output at the same bin, and $m_{n,p} \in \{0,1\}$ is a binary mask (1 = this bin is a peak; 0 = not peak).

\section{Experimental Setup}

\subsection{Dataset}
We use six AIMD RDF dataset taken from \citep{senjaya2020} for amorphous $\text{Ge}_{0.25}\text{Se}_{0.75}$ as binary and  $\text{Ag}_{x}(\text{Ge}_{0.25}\text{Se}_{0.75})_{100-x}$ with $x=(5,10,15,20,25)$ as ternary A to E materials, respectively. Specifically, the RDF simulation data were obtained from Ab Initio Molecular Dynamics (AIMD) using the SIESTA software, where each dataset contains 58 pairs. Meanwhile, the data of the relationship between $q$ and $S_R$($q$) (reduced structure factor) with 148 pairs for all amorphous materials was obtained from the X-ray diffraction experiment using an X-ray source of Cu K-$\alpha$ with $\lambda = 1.54 \ \text{\AA}$ wavelength, also taken from \citep{senjaya2020}. The X-ray diffraction produced a $q_{\text{min}}$ value of 0.65 $\text{\AA}^{-1}$ and a $q_{\text{max}}$ value of 8.00 $\text{\AA}^{-1}$, where these values correspond to scattering angles of 4.57\degree and 78.8\degree. It can be seen that the main peak locations of $S_{R}(q)$ are consistent with the $S(q)$ simulation results from \citep{aimd} in $q_{peak}=\lbrace{1.05,2.05, 3.55, 5.70\rbrace}$. Therefore, the used $S_{R}(q)$ data in this study is considered valid for input data. The data on the relationship between $q$ and $S_R$($q$) for all amorphous materials were corrected from background scattering, normalization, and correction steps, using the method described in \citep{Petkov1989}.

\subsection{Training implementation}
All models are trained with a Graphics Processing Unit (GPU) and implemented with PyTorch. The WT-RDF$+$ (100 epochs) and Machine Learning (ML) models (500 epochs) are trained from scratch with the Adam optimizer \citep{adam} with a learning rate of 0.01. The ML models are pre-processed by a standard scaler or Z-score normalization \citep{normalization} to stabilize the training. For WT-RDF$+$, we trained the model in two rounds. After finishing the first round of training, we replace the original parameters with the optimized set and modify the parameter bound values. The learning rate is decreased to 0.001 during the second stage of training to search for a better local minimum.

\subsection{Evaluation metrics}
To assess the reconstruction performance, the Mean Absolute Error (MAE), First Peak Error (FPE), and Second Peak Error (SPE) are utilized as quantitative metrics \citep{craven, kobayashi}. A model is considered structurally accurate if peaks are also well-aligned. \citep{kobayashi} demonstrated the sensitivity of the first and second RDF peaks to local and medium-range structure in high-density silica glass. 

\begin{equation}
\label{eq:MAE_loss}
\text{MAE} = \frac{1}{N} \sum_{n=1}^{N} \left| y_n - \hat{y}_n \right|
\end{equation}

\begin{equation}
\label{eq:FPE}
\text{FPE} = m_{n} \left| y_{n} - \hat{y}_{n} \right|
\end{equation}

\begin{equation}
\label{eq:SPE}
\text{SPE} = m_{n} \left| y_{n} - \hat{y}_{n} \right|
\end{equation}

\section{Results}

\begin{figure}[ht]
    \centering
    \subfloat[\centering]{{\includegraphics[width=5cm]{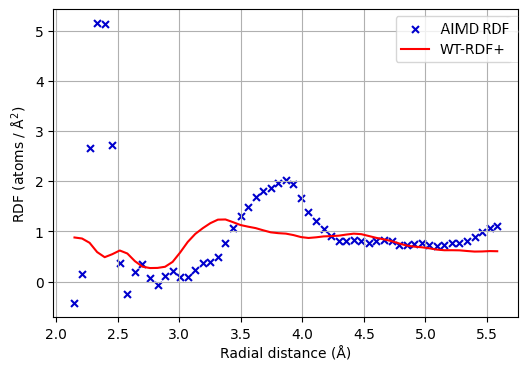}}}
    \hspace{0.1em}
    \subfloat[\centering]{{\includegraphics[width=5cm]{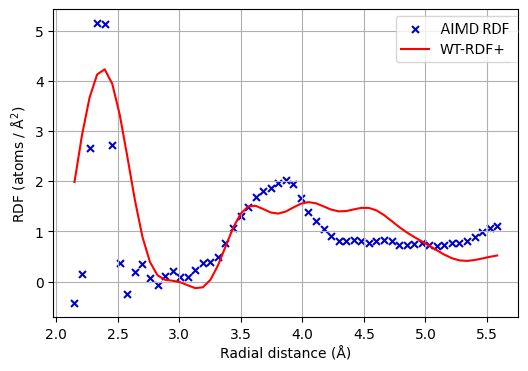} }}
    \hspace{0.1em}
    \subfloat[\centering]{{\includegraphics[width=5cm]{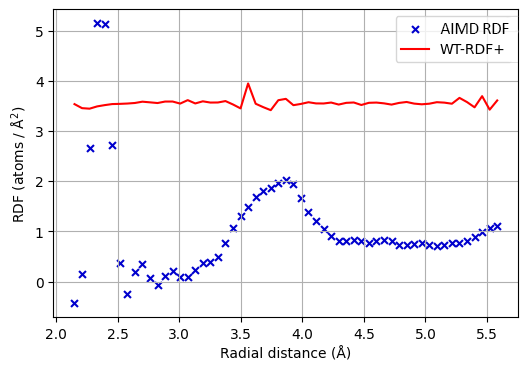} }}
    \hspace{0.1em}
    \subfloat[\centering]{{\includegraphics[width=5cm]{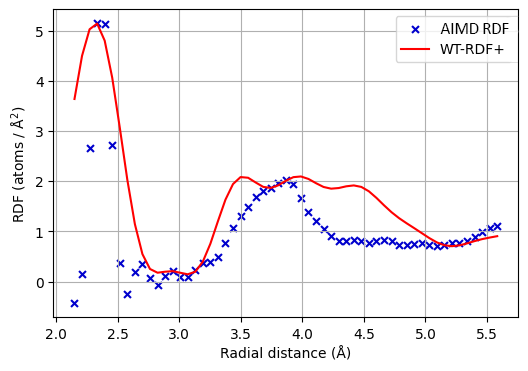} }}
    \caption{The effects of WT-RDF training components on binary data. (a) Parameter optimization without parameter bounding and selective loss; (b) Parameter optimization with parameter bounding; (c) Parameter optimization with selective loss; (d) The result of our proposed WT-RDF${+}$.}
    \label{fig:ablation}
\end{figure}

\begin{table*}[ht]
\renewcommand{\arraystretch}{1.4}
\caption{Ablation study on binary data. MAE, FPE, and SPE are used as metrics. The lower the better. The best results are highlighted in \textbf{bold}. Methods in (a)-(d) are individually demonstrated in Fig. \ref{fig:ablation}.}
\label{table_ablation_study}
\resizebox{\textwidth}{!}{
\centering
\begin{tabular}{c c c c c c c c}
\hline
\bfseries Method
& \bfseries Optimization
& \bfseries Param. Bounding
& \bfseries $\mathcal{L}_{SL}$
& \bfseries MAE $\downarrow$
& \bfseries FPE $\downarrow$
& \bfseries SPE $\downarrow$
& \bfseries Parameter
\\
\hline
&  &  &  & 0.9463 & \textbf{0.0057} & 0.3207 & $a=0.590, b=4.600, K_{f}=\ \ 0.106, \tilde{C}=0.212, \Lambda=-4.700$
\\
(a) & \cmark &  &  & 0.5815 & 4.6346 & 1.0722 & $a=0.348, b=4.587, K_{f}=-0.082, \tilde{C}=0.080, \Lambda=-4.582$
\\
(b) & \cmark & \cmark &  & \textbf{0.5770} & 0.8944 & 0.6312 & $a=0.620, b=4.635, K_{f}=-0.035, \tilde{C}=0.085, \Lambda=-4.405$
\\
(c) & \cmark &  & \cmark & 2.6627 & 1.6005 & 1.6166 & $a=0.014, b=4.596, K_{f}=-0.057, \tilde{C}=0.712, \Lambda=-5.025$
\\
(d) & \cmark & \cmark & \cmark & 0.6598 & 0.0065
 & \textbf{0.0199} & $a=0.610, b=4.471, K_{f}=\ \ 0.010, \tilde{C}=0.133, \Lambda=-5.066$
\\

\hline
\end{tabular}}
\end{table*}

\subsection{Ablation study}
To verify the effects of parameter tuning, parameter bounding, and selective loss, an ablation analysis is performed by omitting each component. The complemented results (a)-(d) are shown in Table \ref{table_ablation_study} and Fig. \ref{fig:ablation}. We observe that training the parameters on Wavelet Transform Radial Distribution Function (WT-RDF) alone can indeed lower the Mean Absolute Error (MAE) on average data by 38.78\%. However, in Fig. \ref{fig:ablation} (a), the first and second peaks are not attained in this case. By bounding a specific sensitive parameter, the MAE, First Peak Error (FPE), and Second Peak Error (SPE) drastically improve. Since WT-RDF is a physics-based model, the radial function is naturally present. In this case, the MAE loss is replaced with a selective loss, allowing the model to focus on the first and second peaks in the data. Furthermore, employing the selective loss without parameter bounding results in a worsening of the outcome because the sensitive parameter matters, as shown in Fig. \ref{fig:ablation} (c). Specifically, when parameter $a$ is not constrained, it continues to decrease and drastically alter parameter $\tilde{C}$ in order to compensate for the impact, which worsens the results. It can also be seen that the selective loss causes the reconstruction to lie between the first and second peaks, meaning that the learning outcome aligns with the intended objectives. By utilizing all components, we can achieve the best result and obtain the trained parameters, improving the WT-RDF to WT-RDF$+$.

\begin{figure}[ht]
    \centering
    \subfloat[\centering]{{\includegraphics[width=5cm]{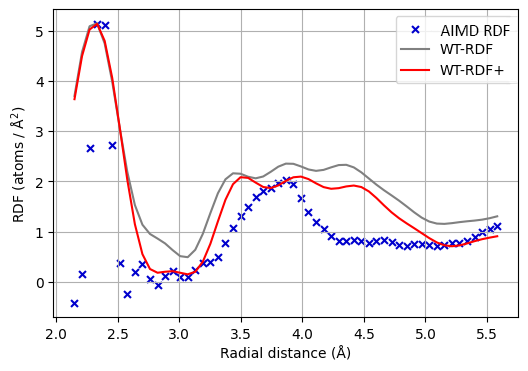}}}
    \hspace{0.1em}
    \subfloat[\centering]{{\includegraphics[width=5cm]
    {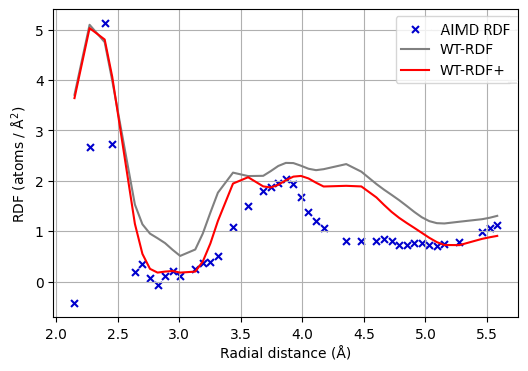} }}
    \hspace{0.1em}
    \subfloat[\centering]{{\includegraphics[width=5cm]
    {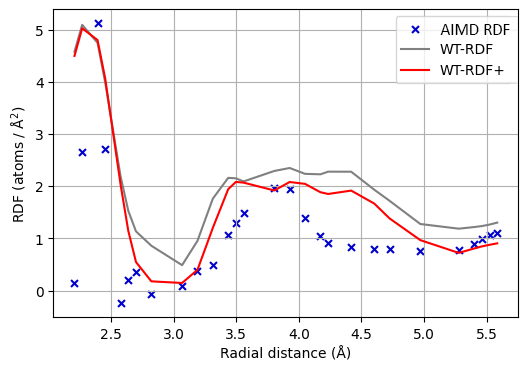} }}
    \hspace{0.1em}
    \subfloat[\centering]{{\includegraphics[width=5cm]
    {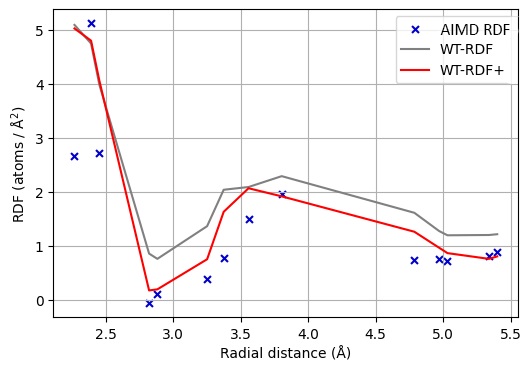} }}
    \caption{Comparison between WT-RDF (baseline) and our proposed WT-RDF${+}$ under different training data ratio on binary data. (a) 100$\%$; (b) 75$\%$; (c) 50$\%$; (d) 25$\%$.}
    \label{fig:wtrdf_scaling}
\end{figure}

\begin{table}[ht]
\renewcommand{\arraystretch}{1.3}
\caption{Comparison of WT-RDF and WT-RDF$+$ with different data ratios on binary data. MAE, FPE, and SPE are used as metrics. The lower the better. The best results are highlighted in \textbf{bold}.}
\label{table_datascale}
\centering
\begin{tabular}{c l c c c}
\hline
\bfseries Data Ratio
& \bfseries Method
& \bfseries MAE $\downarrow$
& \bfseries FPE $\downarrow$
& \bfseries SPE $\downarrow$
\\
\hline
\multirow{2}{*}{100\%}  & WT-RDF & 0.9463 & \textbf{0.0057} & 0.3207
\\
& WT-RDF$+$ & \textbf{0.6598} & 0.0065 & \textbf{0.0199}
\\
\hline
\multirow{2}{*}{75\%}  & WT-RDF & 0.8327 & 0.3750 & 0.3245
\\
& WT-RDF$+$ & \textbf{0.5234} & \textbf{0.3224} & \textbf{0.0199}
\\
\hline
\multirow{2}{*}{50\%}  & WT-RDF & 1.0101 & 0.3750 & 0.3334
\\
& WT-RDF$+$ & \textbf{0.7528} & \textbf{0.3224} & \textbf{0.0406}
\\
\hline
\multirow{2}{*}{25\%}  & WT-RDF & 0.8134 & 0.3750 & 0.3334
\\
& WT-RDF$+$ & \textbf{0.5147} & \textbf{0.3224} & \textbf{0.0406}
\\

\hline
\end{tabular}
\end{table}

\subsection{Data scaling on \texorpdfstring{WT-RDF$+$}{WT-RDF+}}
To test the generality performance of WT-RDF$+$ compared with WT-RDF, we present the results in Table \ref{table_datascale} and Fig. \ref{fig:wtrdf_scaling}. Specifically, we also compare under different data ratios to validate the models' ability to interpolate data. We observe that the proposed WT-RDF$+$ outperforms WT-RDF in terms of MAE, FPE, and SPE. This is further supported by the plotting visualization of the results.

\begin{table}[!ht]
\renewcommand{\arraystretch}{1.3}
\caption{Comparison of WT-RDF$+$ and ML approaches with different data ratios on binary data. MAE, FPE, and SPE are used as metrics. The lower the better. The best results are highlighted in \textbf{bold}. The \textcolor{red}{red} text indicates high error \textgreater 1.}
\label{table_ml_datascale}
\centering
\begin{tabular}{c c c c c}
\hline
\bfseries Data Ratio
& \bfseries Method
& \bfseries MAE $\downarrow$
& \bfseries FPE $\downarrow$
& \bfseries SPE $\downarrow$
\\
\hline
\multirow{3}{*}{100\%} & RBF & \textbf{0.0286} & 0.0340 & 0.0331
\\
& LSTM & 0.0229 & 0.0370 & 0.0377
\\
& WT-RDF$+$ & 0.6598 & \textbf{0.0065} & \textbf{0.0199}
\\
\hline
\multirow{3}{*}{75\%} & RBF & \textbf{0.3223} & \textcolor{red}{1.1011} & 0.6140 
\\
& LSTM & 0.4656 & 0.0769 & \textcolor{red}{1.2753}
\\
& WT-RDF$+$ & 0.7803 & \textbf{0.0042} & \textbf{0.1646}
\\
\hline
\multirow{3}{*}{50\%} & RBF & \textbf{0.5434} & \textcolor{red}{5.5705} & 0.2233
\\
& LSTM & 0.7193 & \textcolor{red}{2.3270} & 0.6587
\\
& WT-RDF$+$ & 0.7468 & \textbf{0.1224} & \textbf{0.0194}
\\
\hline
\multirow{3}{*}{25\%} & RBF & 0.8522 & \textcolor{red}{4.9250} & \textcolor{red}{1.2345}
\\
& LSTM & 0.8365 & \textcolor{red}{5.3931} & 0.6486
\\
& WT-RDF$+$ & \textbf{0.7296} & \textbf{0.0620} & \textbf{0.0444}
\\
\hline
\end{tabular}
\end{table}

\begin{figure}
    \centering
    \subfloat[\centering]{{\includegraphics[width=5cm]{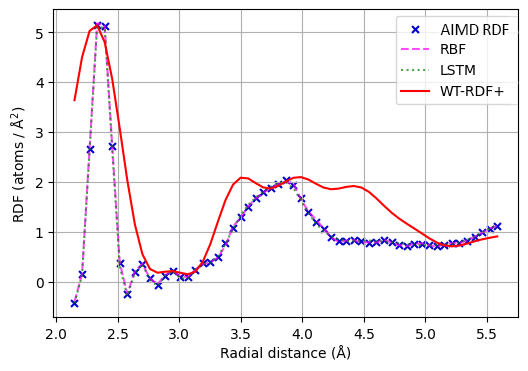}}}
    \hspace{0.1em}
    \subfloat[\centering]{{\includegraphics[width=5cm]{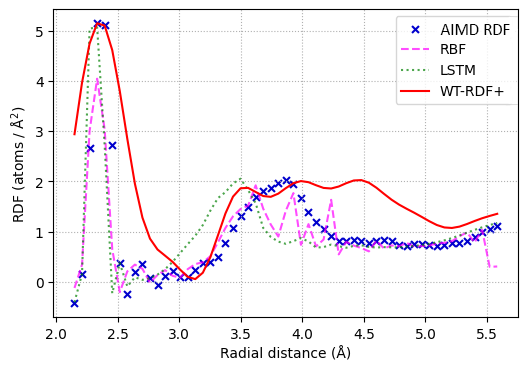} }}
    \hspace{0.1em}
    \subfloat[\centering]{{\includegraphics[width=5cm]{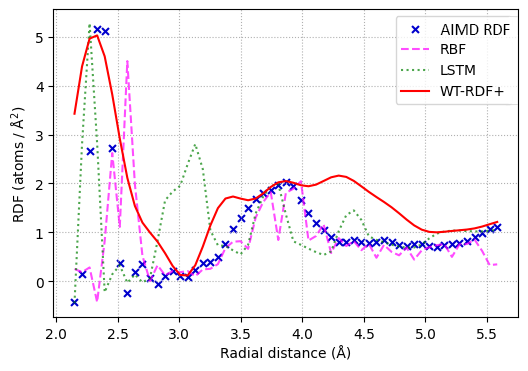} }}
    \hspace{0.1em}
    \subfloat[\centering]{{\includegraphics[width=5cm]{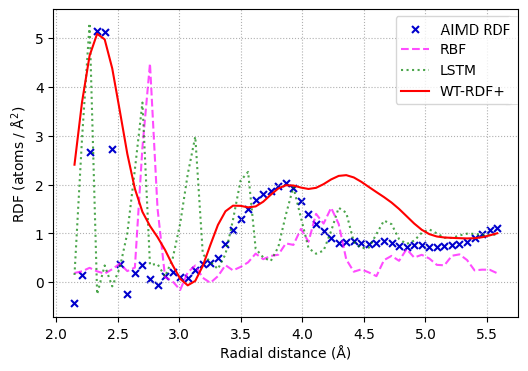} }}
    \caption{Comparison between ML approaches and our proposed WT-RDF${+}$ under different training data ratio on binary data. (a) 100$\%$; (b) 75$\%$; (c) 50$\%$; (d) 25$\%$.}
    \label{fig:ml_scaling}
\end{figure}

\subsection{Comparison with Machine Learning}
We compare our physics-based model with Machine Learning (ML) models shown in Table \ref{table_ml_datascale} and Fig. \ref{fig:ml_scaling}. It has been acknowledged that the data of amorphous materials is generally limited. Additionally, the ML method is data-driven, relying heavily on the data. To test this statement, WT-RDF$+$ and ML models are first trained on different data ratios, \textit{i.e.}, 100$\%$, 75$\%$, 50$\%$, and 25$\%$. Specifically, a portion of the data is randomly selected to ensure its representativeness. Subsequently, WT-RDF$+$ and ML models trained on a smaller portion of the datasets are used to reconstruct the full data (100$\%$). We observe that (1) ML approaches are inherently data-driven, relying on patterns learned from the training data. However, it is noted that 100\% training data with the same dataset for evaluation presented in Fig. \ref{fig:ml_scaling} (a) shows a demonstration of how ML models overfit, causing a prediction bias. This is why the MAE of ML models in Table \ref{table_ml_datascale} is low. Therefore, to verify the ML model's performance, the model must also reconstruct unseen data, as shown in Fig. \ref{fig:ml_scaling} (b-d). When the training data is inadequate, the model is unlikely to perform well on the full dataset.  (2) Based on fundamental mathematical principles, the physics-based model, WT-RDF$+$, maintains reliable performance under limited-data conditions, concretely using only 25\% of the available data. These experimental results verify that our WT-RDF$+$ is a solid reconstructive model alternative for amorphous data.

\begin{figure}
    \centering
    \subfloat[\centering]{{\includegraphics[width=5cm]{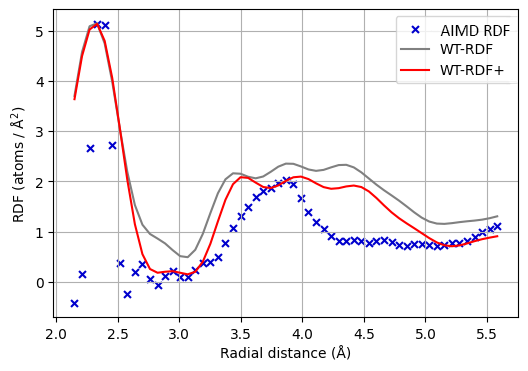}}}
    \subfloat[\centering]{{\includegraphics[width=5cm]{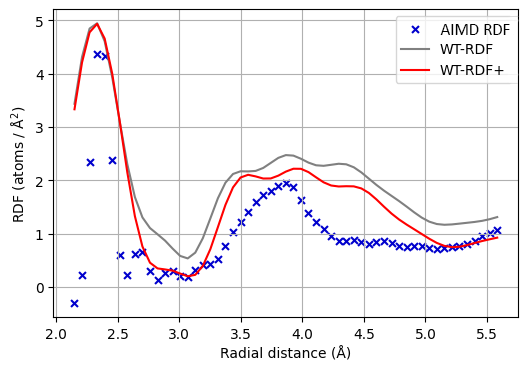} }}
    \hspace{0.1em}
    \subfloat[\centering]{{\includegraphics[width=5cm]{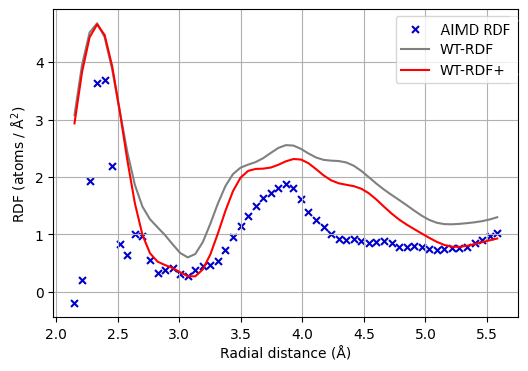} }}
    \subfloat[\centering]{{\includegraphics[width=5cm]{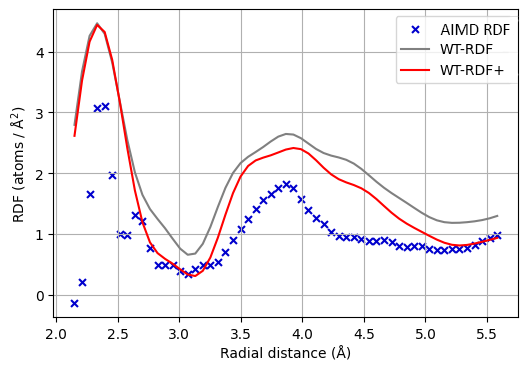} }}
    \hspace{0.1em}
    \subfloat[\centering]{{\includegraphics[width=5cm]{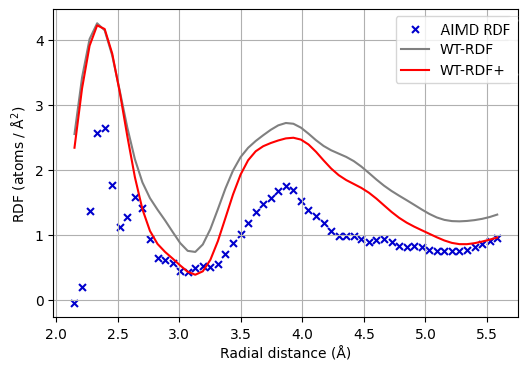} }}
    \subfloat[\centering]{{\includegraphics[width=5cm]{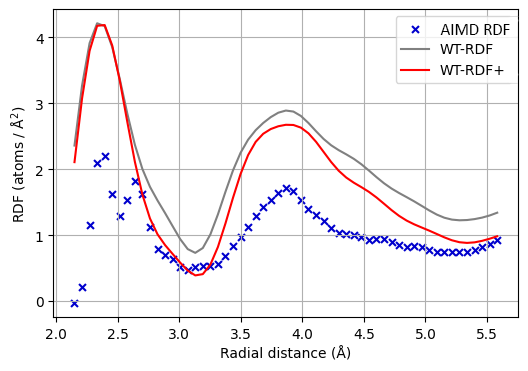} }}
    \caption{Performance comparison between WT-RDF and WT-RDF$+$ trained on binary data across multiple datasets. (a) binary; (b) ternary A; (c) ternary B; (d) ternary C; (e) ternary D; (f) ternary E;}
    \label{fig:cross_dataset}
\end{figure}

\subsection{Cross-dataset evaluation}
To evaluate cross-dataset generalization, we used the optimized parameter of WT-RDF$+$ from the binary data. The models are then evaluated on Ag-doped Ge-Se (ternary A, B, C, D, and E). In Fig. \ref{fig:cross_dataset}, a comparison between WT-RDF$+$ and WT-RDF reveals that WT-RDF$+$ consistently delivers improved performance. The optimization improves the adaptability of WT-RDF$+$ to diverse data distributions while maintaining physical consistency. By balancing the physical prior with machine learning tuning, WT-RDF$+$ captures subtle variations across datasets, resulting in improved reconstructive performance on Ag-doped Ge-Se datasets compared to WT-RDF.

\begin{figure}
    \centering
    \subfloat[\centering]{{\includegraphics[width=5cm]{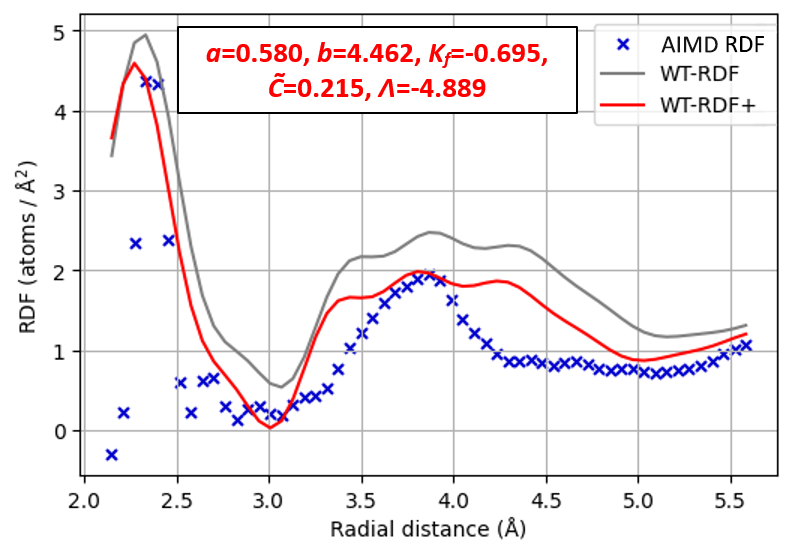}}}
    \subfloat[\centering]{{\includegraphics[width=5cm]{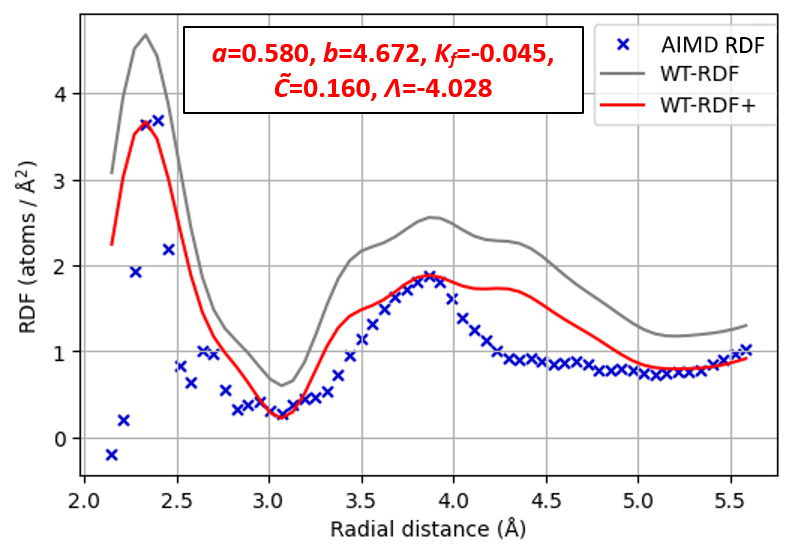} }}
    \hspace{0.1em}
    \subfloat[\centering]{{\includegraphics[width=5cm]{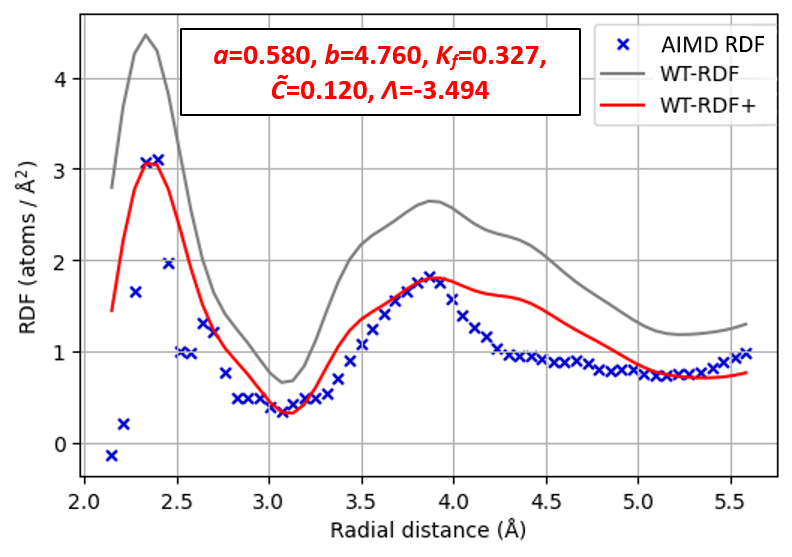} }}
    \hspace{0.1em}
    \subfloat[\centering]{{\includegraphics[width=5cm]{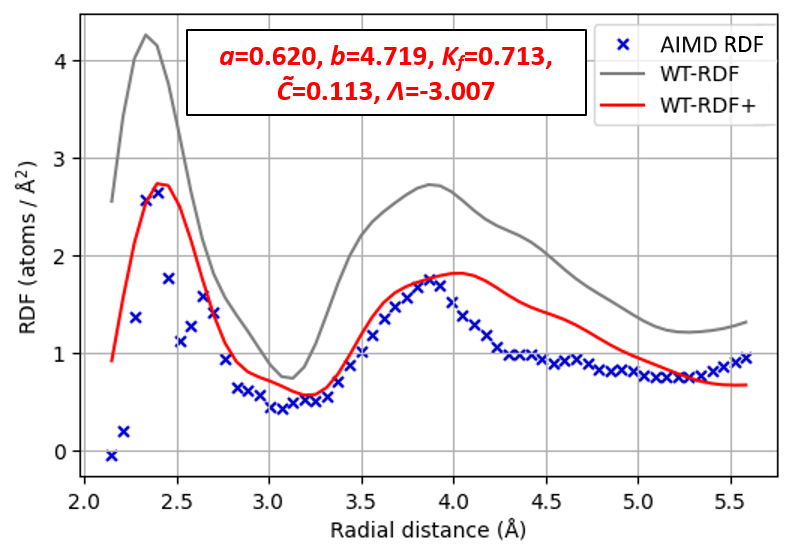} }}
    \subfloat[\centering]{{\includegraphics[width=5cm]{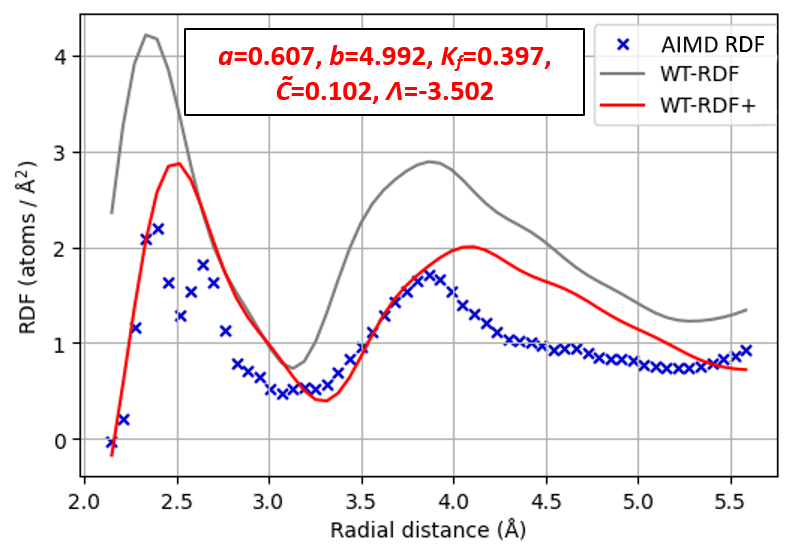} }}
    \caption{Comparison between WT-RDF and WT-RDF$+$ in parameter tuning across different datasets. (a) ternary A; (b) ternary B; (c) ternary C; (d) ternary D; (e) ternary E;}
    \label{fig:internal}
\end{figure}

\subsection{Optimization evaluation}

In Fig. \ref{fig:internal}, the proposed tuning strategy exhibits robust effectiveness on multiple datasets. We also report the parameter changes resulting from the optimization. On the basis of the results, combining physics and machine learning yields improved reconstructions through mutual synergy. This demonstrates the consistency of reconstructive performance across all datasets. In addition, it is worth mentioning that, in this study, we present the optimal parameters deduced from training on binary data ($\text{Ge}_{0.25}\text{Se}_{0.75}$), as it represents stable and simpler amorphous materials \citep{francesco2020, sava2021, Boukhvalov2021}.

\begin{table}
\renewcommand{\arraystretch}{1.3}
\caption{Variability in the performance of WT-RDF$+$.}
\label{table_std}
\centering
\resizebox{\textwidth}{!}{
\begin{tabular}{c c c c }
\hline
\bfseries MAE $\downarrow$
& \bfseries FPE $\downarrow$
& \bfseries SPE $\downarrow$
& \bfseries Parameter
\\
\hline
0.6598 $\pm$ 0.0248 & 0.0065 $\pm$ 0.0012 & 0.0199 $\pm$ 0.0038 & $a=0.610 \pm 0.0058, b=4.471 \pm 0.1443, K_{f}=0.010 \pm 0, \tilde{C}=0.133 \pm 0.0363, \Lambda=-5.066 \pm 0.4380$
\\
\hline
\end{tabular}
}
\end{table}

\subsection{Performance and parameter variability}

As training may involve random elements, such as weight initialization, the results can vary with each run. Our models are trained over three runs. In Table \ref{table_std}, we report the errors and parameters of our optimized model, along with the standard deviation to assess the variability. We observe that the parameter values obtained from machine learning assistance are sensible according to the characteristics of the parameters discussed in Section 3.2.1.

\begin{figure}[H]
    \centering
    \includegraphics[scale=0.43]{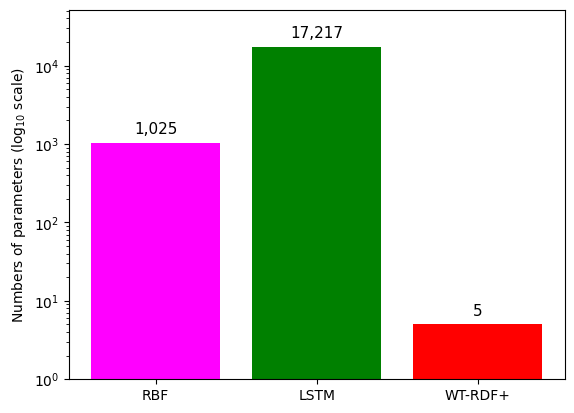}
    \caption{Numbers of parameter comparison of WT-RDF$+$ and ML methods.}
    \label{fig:parameter_efficiency}
\end{figure}

\subsection{Parameter efficiency}

We evaluate the parameter efficiency of each method. Fig. \ref{fig:parameter_efficiency} shows that our WT-RDF$+$ only has 5 parameters compared to other ML methods, \textit{i.e.}, Radial Basis Function (RBF) and Long Short-term Memory (LSTM). Specifically, we merely update the parameters $a$, $b$, $K_f$, $\tilde{C}$, and $\Lambda$. This result suggests the parameter efficiency of our physics-based model. WT-RDF$+$ achieves stable learning with significantly fewer samples due to its compact parameterization and strong physical inductive biases. This is particularly advantageous in industrial problems where data acquisition is expensive and physically constrained.

\begin{table}[h]
\renewcommand{\arraystretch}{1.3}
\caption{Comparison of optimization algorithms.}
\label{table_opt}
\centering
\begin{tabular}{c c c c }
\hline
\bfseries Method
& \bfseries MAE $\downarrow$
& \bfseries FPE $\downarrow$
& \bfseries SPE $\downarrow$
\\
\hline
L-BFGS-B (w/o $\mathcal{L}_{SL}$) & \textbf{0.5092} & 1.3283 & 0.5398
\\
L-BFGS-B (w/ $\mathcal{L}_{SL}$) & 0.7015 & 0.2166 & 0.0243
\\
Adam \citep{adam} & 0.6598 &	\textbf{0.0065} & \textbf{0.0199}
\\
\hline
\end{tabular}
\end{table}

\subsection{Optimization algorithm}
{
As shown in Table \ref{table_opt}, we compare the used optimization algorithm (Adam \citep{adam}) in this paper with the classical one, i.e., L-BFGS-B. L-BFGS-B \citep{lbfgs} is an algorithm solving large nonlinear optimization problems with bounding variable. In this study, we use the same training settings as our approach to train with the L-BFGS-B including the parameter bounding values of $a$ and $K_f$, training iterations, two-rounds training, and the use of selective loss function for a fair comparison. For training, the initial parameters are $a=0.590, b=4.600, K_{f}=0.106, \tilde{C}=0.212, \Lambda=-4.700$. On the basis of the result, Adam achieves better performance compared to L-BFGS-B. We conjecture that the $S_{R}(q)$ data is noisy and non-convex. Therefore, stochastic method like Adam is designed to handle such issue as adaptive learning rates and momentum are incorporated, navigating noisy and non-convex landscapes more effectively. It has also been demonstrated that selective loss plays a vital role in enhancing the FPE and SPE results.
}

\section{Discussion}
\subsection{Why Investigating Atomic Structure is Important?}
A material consists of atoms bonded together, and these interatomic bonds are accurately described using Quantum Mechanics (QM) through the Schrödinger equation \citep{mqc,mmqm,brian2012} shown in Eq. (\ref{discuss1}).

\begin{equation}
    \label{discuss1} \hat{H}\Psi(\vec{R}_{1}...\vec{R}_{i},\vec{r}_{1}...\vec{r}_{j})=E\Psi(\vec{R}_{1}...\vec{R}_{i},\vec{r}_{1}...\vec{r}_{j})
\end{equation}

Where $\Psi(\vec{R}_{1}...\vec{R}_{i},\vec{r}_{1}...\vec{r}_{j})$ is the system’s wave function describing the interatomic bonds. The Hamiltonian operator $\hat{H}$ for a many-atom system, such as a material, can be written as in Eq. (\ref{discuss2}) \citep{mqc,mmqm}.

\begin{equation}
\label{discuss2}
    \hat{H} = \hat{T}_{n}+\hat{T}_{e}+\hat{V}_{en}+\hat{V}_{ee}+\hat{V}_{nn}
\end{equation}
Here, $\hat{T}_{n}$ and $\hat{T}_{e}$ are the kinetic energy operators of the nuclei and electrons (Eq. (\ref{discuss3}) and Eq. (\ref{discuss4})), while $\hat{V}_{en}$, $\hat{V}_{ee}$, and $\hat{V}_{nn}$ represent the electron–nuclei, electron–electron, and nuclei–nuclei interaction potentials (Eq. (\ref{discuss5}), Eq. (\ref{discuss6}), and Eq. (\ref{discuss7})).   

\begin{equation}
\label{discuss3}
    \hat{T}_{n}=\sum_{i}\frac{-\hbar^{2}}{2m_{i}}\nabla_{i}^{2}
\end{equation}
\begin{equation}
\label{discuss4}
    \hat{T}_{e}=\frac{-\hbar^{2}}{2m_{e}}\sum_{j}\nabla_{j}^{2}
\end{equation}
\begin{equation}
\label{discuss5}
    \hat{V}_{en}=-\frac{1}{4\pi\epsilon_{o}}\sum_{ij}\frac{Z_{i}e^{2}}{|\vec{R}_{i}-\vec{r}_{j}|}
\end{equation}
\begin{equation}
\label{discuss6}
    \hat{V}_{ee}=\frac{e^{2}}{4\pi\epsilon_{o}}\sum_{j,k>j}\frac{1}{|\vec{r}_{j}-\vec{r}_{k}|}
\end{equation}
\begin{equation}
\label{discuss7}
    \hat{V}_{nn}=\frac{e^{2}}{4\pi\epsilon_{o}}\sum_{i,l>i}\frac{Z_{i}Z_{l}}{|\vec{R}_{i}-\vec{R}_{j}|}
\end{equation}
Here, $\epsilon_{0}$ is the vacuum permittivity ($8.85\times10^{-12} \ \text{C}^2/\text{N}\text{m}^2)$, $m_{i}$ is the mass of the \textit{i}-th atomic nucleus, and $m_{e}=9.1\times10^{-31} \ \text{kg}$ is the electron mass. Meanwhile, $Z_{i}$ denotes the atomic number of each atom in the system.

Electrons have a much smaller mass than atomic nuclei ($m_{e}\approx10^{-31} \ \text{kg} \ll m_{i}\approx10^{-27} \ \text{kg}$) \citep{born1927}. As a result, electrons respond far more rapidly and collectively to interactions, allowing the kinetic energy of the nuclei to be neglected and the nuclei–nuclei potential to be treated as constant. These simplifications reduce Eq. (\ref{discuss2}) to the simpler form in Eq. (\ref{discuss8}). This reduction is known as the Born–Oppenheimer approximation \citep{born1927}.

\begin{equation}
\label{discuss8}
    \hat{H}\approx \hat{T}_{e}+\hat{V}_{ee}+\hat{V}_{en} \rightarrow \hat{H}\Psi(\vec{r}_{1}...\vec{r}_{j})=\bar{E}\Psi(\vec{r}_{1}...\vec{r}_{j})
\end{equation}
The approach in Eq. (\ref{discuss8}) shows that a material’s properties are largely governed by the behavior and distribution of its electrons, which are closely linked to how the atoms are arranged. Thus, Eq. (\ref{discuss8}) underscores the importance of understanding a material’s underlying atomic structure.

\subsection{Why Physics-based model?}
In experimental practice, Radial Distribution Function (RDF) determination often begins with X-ray diffraction to find the expression $S_{R}(q)$ and transform the data based on Fourier Transform (FT). However, the use of X-ray diffraction often has limitations. Two examples of these limitations are related to the detectable angle range and the resolution of the X-ray diffraction instrument. The angle range directly affects the wave vector $q$, resulting in minimal data availability for FT input. Furthermore, the poor resolution of X-ray diffraction instruments affects the pre-data for extracting $S_{R}(q)$. This, of course, also has a significant impact on the poor resolution of the RDF produced through FT.

Based on these two facts, a physics-based RDF model is needed. The physics-based RDF model developed in this study, WT-RDF+, enables us to reconstruct RDF with good resolution, despite minimal data availability and the resolution of the X-ray diffraction instrument itself. The reason is that through physics, WT-RDF+ has been able to provide an overview of the typical main interactions between atoms and corrections to the interaction terms in the design of its wavelet function. In addition, WT enables the ability to act as an extraordinary mathematical microscope for function analysis on a small (local) scale, as explained in Section 2. In addition to being physics-based, WT-RDF+ excels because its parameter determination is assisted by the backpropagation algorithm of Machine Learning (ML). The ML assistance presented here is intended to guide the determination of model parameters that are more accurate and enhance the reconstructive capabilities of the existing WT-RDF, rather than building an ML model or a hybrid model.

\section{Conclusion}
This research significantly enhances the accuracy and reconstructive performance of the Wavelet Transform Radial Distribution Function (WT-RDF) model by Machine Learning method via learnable parameter ($a$, $b$, $K_{f}$, $\tilde{C}$, and $\Lambda$) optimization, parameter bounding, and selective loss, resulting in the improved WT-RDF+. Its effectiveness is demonstrated by more precise reconstructions of the first and second RDF peaks and the overall RDF trends in binary $\text{Ge}_{0.25}\text{Se}_{0.75}$ (binary) and $\text{Ag}_{x}(\text{Ge}_{0.25}\text{Se}_{0.75})$$_{100-x}$ $(x=5,10,15,20,25)$ amorphous systems by requiring a reference $G(r)$ (AIMD data) for calibration. WT-RDF+ also outperforms Machine Learning models such as Radial Distribution Function (RBF) and Long Short-term Memory (LSTM), especially with limited data. With only 25\% of the training data, it reduces the FPE by 98.74\% vs. RBF and 98.85\% vs. LSTM, and the SPE by 96.40\% and 93.15\%, respectively. These results highlight WT-RDF+ as a robust tool for advancing technology and manufacturing with Ge-Se and Ag-Ge-Se materials.

\newpage

\bibliographystyle{elsarticle-num}
\bibliography{reference}

\newpage

\setcounter{table}{0}
\renewcommand{\thetable}{A\arabic{table}}

\setcounter{figure}{0}
\renewcommand{\thefigure}{A\arabic{figure}}

\appendices

\begin{table}[htbp]
\centering
\caption{Summary and sample data points of the binary dataset. $S_{R}(q) = q(S(q)-1)$}
\label{tab:dataset_samples}
\renewcommand{\arraystretch}{1.2}
\resizebox{\columnwidth}{!}{
\begin{tabular}{lccl}
\hline
\textbf{Dataset} & \textbf{Total pairs} & \textbf{Data} & \textbf{Sample (10 points per row)} \\
\hline
RDF Exp. 
& 58 
& r 
& 2.08621, 2.14757, 2.20893, 2.27029, 2.33165, 2.39301, 2.45437, 2.51573, 2.57709, 2.63845,
\\

& 
& 
& 2.69981, 2.76117, 2.82252, 2.88388, 2.94524, 3.0066, 3.06796, 3.12932, 3.19068, 3.25204, 
\\

&
&
& 3.3134, 3.37476, 3.43612, 3.49748, 3.55884, 3.62019, 3.68155, 3.74291, 3.80427, 3.86563,
\\

&
&
& 3.92699, 3.98835, 4.04971, 4.11107, 4.17243, 4.23379, 4.29515, 4.35651, 4.41786, 4.47922,
\\

&
&
& 4.54058, 4.60194, 4.6633, 4.72466, 4.78602, 4.84738, 4.90874, 4.9701, 5.03146, 5.09282,
\\

&
&
& 5.15418, 5.21553, 5.27689, 5.33825, 5.39961, 5.46097, 5.52233, 5.58369	
\\

& 
& G(r) 
& 0.2251, -0.4308, 0.1542, 2.66594, 5.14732, 5.12395, 2.72029, 0.37955, -0.2425, 0.19537,
\\

& 
& 
& 0.35462, 0.06315, -0.05997, 0.11836, 0.20673, 0.10133, 0.08318, 0.24179, 0.36787, 0.39258,
\\

& 
& 
& 0.49884, 0.77222, 1.07455, 1.30096, 1.49633, 1.68363, 1.80603, 1.87293, 1.96002, 2.02941,
\\

& 
& 
& 1.93852, 1.67098, 1.38718, 1.19846, 1.05383, 0.90346, 0.80962, 0.81383, 0.83634, 0.80566,
\\

& 
& 
& 0.77338, 0.80159, 0.83644, 0.80076, 0.73264, 0.72022, 0.75521, 0.76038, 0.72529, 0.7108,
\\

& 
& 
& 0.73747, 0.76506, 0.77949, 0.81741, 0.89638, 0.98642, 1.05957, 1.11278
\\
\hline

SRQ 
& 148 
& q
& 0.65, 0.7, 0.75, 0.8, 0.85, 0.9, 0.95, 1, 1.05, 1.1,
\\

&
& 
& 1.15, 1.2, 1.25, 1.3, 1.35, 1.4, 1.45, 1.5, 1.55, 1.6,
\\

&
& 
& 1.65, 1.7, 1.75, 1.8, 1.85, 1.9, 1.95, 2, 2.05, 2.1,
\\

&
& 
& 2.15, 2.2, 2.25, 2.3, 2.35, 2.4, 2.45, 2.5, 2.55, 2.6,
\\

&
& 
& 2.65, 2.7, 2.75, 2.8, 2.85, 2.9, 2.95, 3, 3.05, 3.1,
\\

&
& 
& 3.15, 3.2, 3.25, 3.3, 3.35, 3.4, 3.45, 3.5, 3.55, 3.6,
\\

&
& 
& 3.65, 3.7, 3.75, 3.8, 3.85, 3.9, 3.95, 4, 4.05, 4.1,
\\

&
& 
& 4.15, 4.2, 4.25, 4.3, 4.35, 4.4, 4.45, 4.5, 4.55, 4.6,
\\

&
& 
& 4.65, 4.7, 4.75, 4.8, 4.85, 4.9, 4.95, 5, 5.05, 5.1,
\\

&
& 
& 5.15, 5.2, 5.25, 5.3, 5.35, 5.4, 5.45, 5.5, 5.55, 5.6,
\\

&
& 
& 5.65, 5.7, 5.75, 5.8, 5.85, 5.9, 5.95, 6, 6.05, 6.1,
\\

&
& 
& 6.15, 6.2, 6.25, 6.3, 6.35, 6.4, 6.45, 6.5, 6.55, 6.6,
\\

&
& 
& 6.65, 6.7, 6.75, 6.8, 6.85, 6.9, 6.95, 7, 7.05, 7.1,
\\

&
& 
& 7.15, 7.2, 7.25, 7.3, 7.35, 7.4, 7.45, 7.5, 7.55, 7.6,
\\

&
& 
& 7.65, 7.7, 7.75, 7.8, 7.85, 7.9, 7.95, 8
\\

& 
& $S_{R}(q)$
& -0.591721, -0.606298, -0.6297525, -0.64308, -0.6382055, -0.609381, -0.554325, -0.4885, -0.453075, -0.479149,
\\

& 
& 
& -0.5578995, -0.656016, -0.7435875, -0.811655, -0.8614485, -0.89502, -0.9118905, -0.9102, -0.8878555, -0.84096,
\\

& 
& 
& -0.764808, -0.652817, -0.4956, -0.286848, -0.0288045, 0.261801, 0.5452005, 0.75302, 0.831111, 0.774669,
\\

& 
& 
& 0.61705, 0.410498, 0.19683, -0.001725, -0.180245, -0.328992, -0.452662, -0.557975, -0.6429825, -0.711074,
\\

& 
& 
& -0.7613715, -0.794394, -0.809215, -0.805924, -0.781185, -0.727407, -0.6450175, -0.534, -0.3904915, -0.210211,
\\

& 
& 
& 0.0117495, 0.272096, 0.5592275, 0.873015, 1.190858, 1.488894, 1.738662, 1.915025, 1.999431, 1.983456,
\\

& 
& 
& 1.877633, 1.706329, 1.4986875, 1.263424, 1.01255, 0.754923, 0.4906295, 0.22044, -0.045279, -0.295036,
\\

& 
& 
& -0.5368025, -0.76524, -0.971805, -1.148487, -1.293864, -1.406284, -1.485855, -1.53279, -1.54973, -1.538286,
\\

& 
& 
& -1.4990205, -1.435051, -1.3458175, -1.234368, -1.1041025, -0.956529, -0.795861, -0.6223, -0.439956, -0.258468,
\\

& 
& 
& -0.077044, 0.099476, 0.2629725, 0.412817, 0.5494985, 0.67365, 0.790904, 0.89199, 0.9556545, 1.003128,
\\

& 
& 
& 1.0394305, 1.059459, 1.0589775, 1.039708, 1.0000575, 0.93751, 0.8550745, 0.75822, 0.650254, 0.533262,
\\

& 
& 
& 0.4122345, 0.294128, 0.188875, 0.097965, 0.0137795, -0.060352, -0.1119075, -0.14313, -0.156021, -0.154176,
\\

& 
& 
& -0.140581, -0.12194, -0.1044225, -0.087924, -0.0770625, -0.081144, -0.1048755, -0.14749, -0.207834, -0.285207,
\\

& 
& 
& -0.379808, -0.488664, -0.5977625, -0.69715, -0.7947555, -0.874754, -0.9214905, -0.9438, -0.952508, -0.93138,
\\

& 
& 
& -0.876996, -0.795025, -0.693005, -0.573456, -0.439914, -0.297751, -0.1496985, -0.006
\\

\hline
\end{tabular}
}
\end{table}

\begin{algorithm}
\caption{WT-RDF$+$ Training}
\label{alg:wtrdf_compact}
\begin{algorithmic}[1]
\Require Input data $r$, reduced structure factor $S_{R}(q)$
\Ensure Output $\hat{G}(r)$

\State Initialize model parameters $\theta=\{a,b,K_f,\tilde{C},\Lambda\}$
\State Initialize Adam hyperparameters $\alpha,\beta_1,\beta_2,\epsilon$

\State $u \gets 0,\; v \gets 0$

\For{$i = 1$ to $100$}
    \State $\hat{G}(r) \gets f(r,S_{R}(q);\theta)$
    \Comment{Equation 3 in the manuscript}
    \State $\mathcal{L}_{SL} \gets m\|G(r)-\hat{G}(r)\|$
    \Comment{Selective loss from Equation 16}
    \State $g_i \gets \nabla_\theta \mathcal{L}_{SL}$ \Comment{Gradient at epoch $i$}
    \State $u \gets \beta_1 u + (1-\beta_1) g_i$ \Comment{First moment (momentum)}
    \State $v \gets \beta_2 v + (1-\beta_2) g_i^2$ \Comment{Second moment (variance)}
    \State $\hat{u} \gets u/(1-\beta_1^i),\;
           \hat{v} \gets v/(1-\beta_2^i)$
    \Comment{Bias correction}
    \State $\theta \gets \theta - \alpha \hat{u}/(\sqrt{\hat{v}}+\epsilon)$
    \Comment{Parameter update}
\EndFor

\State \Return $\hat{G}(r)$
\end{algorithmic}
\end{algorithm}

To ease a relevant study in the future, a binary dataset $\text{Ge}_{0.25}\text{Se}_{0.75}$ is provided as a sample, as summarized in Table \ref{tab:dataset_samples}. Moreover, the pseudocode of the WT-RDF$+$ training process is also provided in Algorithm \ref{alg:wtrdf_compact} to improve clarity. 

\end{document}